\documentclass[conference]{IEEEtran}

\usepackage{adjustbox}
\usepackage{fancyhdr}
\usepackage{float}
\usepackage[T1]{fontenc}
\usepackage[utf8]{inputenc}
\usepackage{multicol}
\usepackage{multirow}
\usepackage[normalem]{ulem}
\usepackage{tabularx}
\usepackage{amsmath}
\usepackage{graphicx}
\usepackage{algorithm}
\usepackage{algorithmic}
\usepackage{listings} 
\usepackage{xcolor}
\usepackage{url}
\usepackage{tikz}
\usepackage{caption}
\usepackage{PTSansNarrow} 
\usepackage{authblk}
\usepackage{hyperref}
\usepackage{pgfplots}
\usepackage{subcaption}
\usepackage{cite}
\usepackage{amssymb,amsfonts}
\usepackage{textcomp}
\def\BibTeX{{\rm B\kern-.05em{\sc i\kern-.025em b}\kern-.08em
    T\kern-.1667em\lower.7ex\hbox{E}\kern-.125emX}}

\hypersetup{
    hidelinks, 
    pdftitle={Overleaf Example},
    pdfpagemode=FullScreen,
}

\lstdefinelanguage{json}{
    basicstyle=\scriptsize\ttfamily,
    showstringspaces=false,
    breaklines=true,
    frame=lines,
    backgroundcolor=\color{gray!10},
    literate=
     *{0}{{{\color{blue}0}}}{1}
      {1}{{{\color{blue}1}}}{1}
      {2}{{{\color{blue}2}}}{1}
      {3}{{{\color{blue}3}}}{1}
      {4}{{{\color{blue}4}}}{1}
      {5}{{{\color{blue}5}}}{1}
      {6}{{{\color{blue}6}}}{1}
      {7}{{{\color{blue}7}}}{1}
      {8}{{{\color{blue}8}}}{1}
      {9}{{{\color{blue}9}}}{1}
      {:}{{{\color{purple}:}}}{1}
      {,}{{{\color{purple},}}}{1}
      {\{}{{{\color{brown}\{}}}{1}
      {\}}{{{\color{brown}\}}}}{1}
      {[}{{{\color{brown}[}}}{1}
      {]}{{{\color{brown}]}}}{1},
}

\begin{document}

\title{ARLO: A Tailorable Approach for Transforming Natural Language Software Requirements into Architecture using LLMs}
\author{Tooraj Helmi, University of Southern California, thelmi@usc.edu}
\IEEEoverridecommandlockouts
\IEEEaftertitletext{}
\maketitle



\begin{abstract}
Software requirements expressed in natural language (NL) frequently suffer from verbosity, ambiguity, and inconsistency. This creates a range of challenges, including selecting an appropriate architecture for a system and assessing different architectural alternatives. Relying on human expertise to accomplish the task of mapping NL requirements to architecture is time-consuming and error-prone. This paper proposes ARLO, an approach that automates this task by leveraging (1) a set of NL requirements for a system, (2) an existing standard that specifies architecturally relevant software quality attributes, and (3) a readily available Large Language Model (LLM). Specifically, ARLO determines the subset of NL requirements for a given system that is architecturally relevant and maps that subset to a tailorable matrix of architectural choices. ARLO applies integer linear programming on the architectural-choice matrix to determine the optimal architecture for the current requirements. We demonstrate ARLO’s efficacy using a set of real-world examples. We highlight ARLO’s ability (1) to trace the selected architectural choices to the requirements and (2) to isolate NL requirements that exert a particular influence on a system’s architecture. This allows the identification, comparative assessment, and exploration of alternative architectural choices based on the requirements and constraints expressed therein.
\end{abstract}

\section{Introduction}
Software systems are ever-evolving, with increases in size and complexity that call for careful elicitation of requirements and architectural choices~\cite{paul2004}. While it has been long recognized that requirements and architecture co-evolve~\cite{Nuseibeh2001,Spijkman2021}, understanding their interactions, especially early in the software development process, is still an open challenge. More specifically, there is a scarcity of knowledge regarding their alignment, architecture-to-requirements traceability, and conserving architectural knowledge~\cite{cleland2013twin}.

Most software requirements are still captured using natural language (NL)~\cite{shreda2021identifying,dave2022identifying,lago2019architecture}. The informal nature of NLs is a significant obstacle in machine processing of such requirements \cite{6093363}. In the past, researchers have proposed approaches to classify the NL requirements into different categories of functional and non-functional requirements~\cite{shreda2021identifying,dave2022identifying}, to identify quality attributes from NL requirements, to facilitate architectural decisions by leveraging machine learning~\cite{lago2019architecture}, and so on. While promising, these approaches require large, curated datasets and high model-building effort~\cite{esfahani2013guidearch}.

Translating NL requirements to architecture-related design decisions is a cumbersome task that has long been recognized in the research community~\cite{paul2004}. In practice, architects' decisions often rely entirely on their intuition, experience, and deep understanding of the software product \cite{svahnberg2003quality}, \cite{macit2020methods}. While they may have the requisite domain knowledge, architects must manually deal with large numbers of NL requirements that may be ambiguous, incomplete, and internally inconsistent~\cite{dalpiaz2018natural, yadav2021comprehensive}. In addition, the architects' decisions tend to reflect their familiarity with and preference for specific design and/or technological choices. All this frequently leads to sub-optimal choices for the software system under development.

These challenges underscore the importance of an effective methodology for analyzing NL requirements and making corresponding architectural decisions. Recent research on applying LLMs in software engineering, particularly in processing requirements, has shown promise. For example, a study by White et al. \cite{white2023chatgpt} shows how LLMs like ChatGPT improve software engineering tasks such as generating API specifications from requirements lists. Another study by Belzner et al.~\cite{belzner2023large} highlights how LLMs can assist in generating class design and corresponding pseudocode of the important classes and their relationships. While these approaches demonstrate specific applications of LLMs in addressing certain aspects of software engineering, they do not provide a comprehensive solution for using LLMs to make broader architectural decisions.

This paper introduces ARLO, a novel approach for arriving at an effective \underline{A}rchitecture from NL \underline{R}equirements using \underline{L}LMs and \underline{O}ptimization algorithms. The use of LLMs is central to our approach. While current state-of-the-art zero-shot LLMs are not perfect, and the quality of their output can vary, they outperform humans when processing large text corpora. On the other hand, while LLMs excel in processing textual information, they do not possess the expert understanding that a human may have in deriving informed architectural decisions from specific requirements. To address this, we suggest utilizing Quality Attributes (QAs) to bridge the gap between NL requirements and architectural choices. This approach enables an Large Language Models (LLMs) to concentrate on extracting QAs, which we hypothesize the LLM can do effectively. The QAs used in our approach are based on those defined in the ISO/IEC 25010 standard~\cite{ISO-25010:2011}: functional suitability, performance efficiency, compatibility, interaction capability, reliability, security, maintainability, and flexibility. Our approach is motivated by the observation that, while there are potentially countless domain- and application-specific architectural choices in practice, existing zero-shot LLMs can deal much more effectively with the small number of well-understood QAs. 

ARLO uses QAs to arrive at the most suitable architectural choices. An architectural choice is a distinct design decision made or strategy pursued in the development of a software system, shaping its structure, constituent components, their interactions, and overall behavior. For instance, the deployment choices for a system may be \emph{monolithic}, \emph{microservices}, or \emph{peer-to-peer}.  

We employ a systematic approach to extract QAs and organize them in a way that ARLO can utilize effectively to derive architectural choices. The process begins with using LLMs to identify architecturally significant requirements (ASRs), which are requirements that specify key decisions regarding high-level software architecture. We then extract crucial information for each ASR, including implied QAs and existing conditional statements. Next, we apply a specific algorithm to group similar conditions into Condition Groups (CGs). Subsequently, we use LLMs to identify Concurrent Condition Groups (CCGs), which are CGs that can be true simultaneously. ARLO will then suggests distinct architectural choices for each CCG. 

ARLO relies on a matrix that captures architectural knowledge to select architectural choices based on the desired QAs. Values in the matrix indicate an architect's assessment of the impact of each choice on a QA. For example, selecting the \emph{always-on} data caching strategy may impose a performance penalty compared to \emph{offline~first}. ARLO's matrix is designed to be simple to create, flexible to include varying architectural needs and preferences, and tailorable to capture new knowledge or system constraints. Considering the desired QAs and their respective weights, calculated based on their frequency of implication in the requirements, ARLO uses the matrix to determine the optimal architectural choices by leveraging an Integer Linear Programming technique. 

We have applied ARLO in three real-world examples, demonstrating 1) its scalability measured as its execution time for software systems with requirements ranging from several hundred to several thousand and 2) its sensitivity to the chosen matrix and specific requirements called architecture influencing requirements (AIRs). An AIR (or a set of AIRs) is an ASR that, when removed, will change the architectural choices selected by ARLO. 

In summary, our contributions are as follows: \begin{enumerate} 
\item A novel approach for automating software architecture decision-making based on arbitrary natural language requirements by leveraging zero-shot LLMs. 
\item Establishing a connection between architectural decisions and requirements through quality attributes while incorporating the architect's insights into how these attributes influence architectural choices. 
\item Introducing the concept of architecturally influencing requirements (AIRs), driving the architectural decisions. 
\item Publicly available implementation of ARLO.
\end{enumerate}

The remainder of the paper is organized as follows. Section~\ref{sec:related-work} delves into existing research on the role of LLMs in requirements engineering and the influence of QAs on software architecture. Section~\ref{sec:approach} presents ARLO's design and implementation. Section~\ref{sec:eval} details its application on three real-world projects. Limitations and threats to validity are discussed in Section~\ref{sec:ttv}. Lastly, conclusions and future directions are outlined in Section \ref{sec:conclusion}.

\section{Related Work}
\label{sec:related-work}
Standards like ISO/IEC 25010 define essential QAs~\cite{ISO-25010:2011}. The influence of QAs on architectural choices~\cite{al2017quality} and software design decisions~\cite{lago2019architecture} is well-documented \cite{marquez2023architectural, bass2003software, svahnberg2002method}. While some RE approaches recognize the importance of QAs and Non-Functional Requirements (NFRs)~\cite{moreira2002crosscutting, domah2015nerv}, they are often implicitly integrated with functional requirements~\cite{borg2003bad, paetsch2003requirements}. Quality concerns like security, performance, and maintainability are key in shaping a system’s architecture and often require trade-offs between functional and quality requirements; for instance, the Twin Peaks model~\cite{cleland2013twin} highlights the iterative relationship between requirements and architecture.

Historically, identifying QAs has been manual and labor-intensive~\cite{rosenhainer2004identifying, kusters1999identifying}. Recent advancements in NLP have enabled automated approaches~\cite{younas2020extraction, ahmed2023nlp, li2021nfrnet}, such as extracting QAs from user stories~\cite{gilson2019extracting} and classifying user reviews from app stores~\cite{lu2017automatic}. The automated extraction and visualization of quality concerns in~\cite{rahimi2014automated, fazelnia2024lessons} uses data mining and ML to identify quality-related requirements from specifications, helping stakeholders understand their impact across projects. 

Research shows that while precise prediction of a system's architecture that achieves desired QAs is challenging~\cite{chung2012non}, assessments of how architectural decisions impact QAs are feasible~\cite{bachmann2005designing}. This can be accomplished via architectural tactics~\cite{bachmann2003moving}, decision maps~\cite{lago2019architecture}, fuzzy mathematics~\cite{esfahani2013guidearch}, and economic modeling~\cite{kazman2001quantifying}. However, these approaches do not prescribe how to arrive at architectural choices.

Design Structure Matrices (DSM) and design rule theory (DRT) are widely used to analyze software modularity and evolution, as demonstrated by MacCormack et al. and Baldwin~\cite{maccormack2012exploring, baldwin2002option}. DSM also supports decision-making and analyzes architectural decay~\cite{lamantia2008analyzing}. Tools like Titan~\cite{xiao2014titan} link software design with quality analysis, while Huynh et al.~\cite{huynh2008automatic} automate UML transformations into decision models, and Cai et al.~\cite{cai2004software} offers a framework for deriving DSMs to understand modularity better. In ARLO, we use a matrix to link QAs to architectural choices to determine the optimum decisions.

Another contribution of ARLO is traceability. Software traceability links artifacts like requirements, code, and tests, ensuring systems meet client specifications. Recent methods improve manual processes by using ML, ontology, and information retrieval (IR), achieving up to 95\% accuracy~\cite{adithya2021ontoreq}. Hybrid approaches combining ML and reasoning outperform traditional IR by accounting for artifact relationships~\cite{li2020combining}. ARLO automates traceability by mapping architecturally significant and influential requirements to QAs using LLMs and optimization, bypassing static associations via textual similarity used in earlier methods.

Recent work has endeavored to use LLMs to derive software architecture aspects. Eisenreich et al.~\cite{eisenreich2024requirements} propose a method to generate a system's architecture semi-automatically based on requirements using modern AI techniques using  manual iterations, with the architect using prompts to tell the tooling what aspects of the architecture candidate need to be changed. \cite{rukmono2024deductive} recover software architecture from code units using deductive reasoning supported by LLMs. Karetnikov et al.~\cite{karetnikov2024semantic} use knowledge graphs for analyzing architecture decision records, allowing them to be reused on a cross-project basis. Soliman et al.~\cite{soliman2021exploring} explore manual retrieval of architectural knowledge from web search engines, allowing architects to make well-founded design decisions. ARLO enhances existing approaches by automating the extraction of ASRs and mapping them to QAs using LLMs. This automation streamlines a traditionally labor-intensive process.
 
\begin{figure}[t!]
\centering
\includegraphics[width=0.48\textwidth]{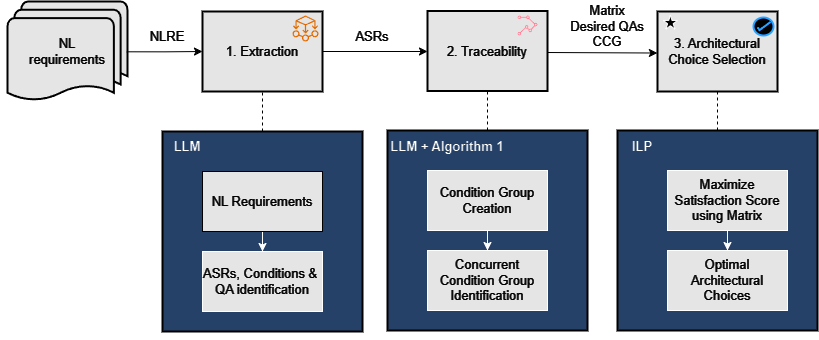}
\caption{Overview of ARLO}
\label{fig:approach}
\end{figure}

\section{The ARLO Approach}
\label{sec:approach}
ARLO automates the selection of an appropriate software {\textbf{\uline{A}}}rchitecture from a set of NL {\textbf{\uline{R}}}equirements by leveraging {\textbf{\uline{L}}}LMs and {\textbf{\uline{O}}}ptimization methods. ARLO supports decision-making by providing appropriate recommendations for architectural choices. It provides a process of architecture discovery~\cite{Spijkman2021} by leveraging the power of LLMs to identify architecturally significant requirements (ASRs) and a set of impacted quality attributes (QAs). 

\begin{figure}[b!]
\centering
\includegraphics[width=0.45\textwidth]{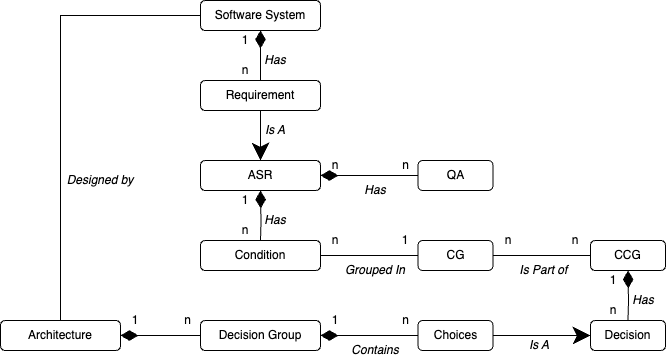}
\caption{ARLO's view of a software system}
\label{fig:concepts}
\end{figure}

ARLO maps the ASRs to corresponding QAs and conditions under which each ASR may occur. Consequently, ARLO generates a collection of concurrent condition groups, each comprising a set of conditions that can simultaneously hold during the system's operation. Lastly, ARLO leverages an off-the-shelf integer linear programming optimizer to select the best architectural choices to recommend to the architect. Figure~\ref{fig:approach} highlights the three key steps involved in ARLO: (1)~extraction of ASRs, (2)~traceability of architectural decisions by grouping desired QAs, and (3)~selecting optimal architectural choices. ARLO also employs various concepts to tackle the complex problem of making architectural choices. Table~\ref{tab:concepts} provides an overview of these key concepts, while Figure~\ref{fig:concepts} presents a UML diagram illustrating the relationships between them.

\begin{table}[htbp]
\centering
\begin{footnotesize}
\caption{Key Concepts in ARLO's Approach}
\label{tab:concepts}
\begin{tabularx}{\columnwidth}{|p{2.5cm}|X|}
\hline
\textbf{Concept} & \textbf{Description} \\ \hline
\textbf{Architecturally Significant Requirement (ASR)} & A requirement that directly impacts the architecture of the system. ASRs imply one or more QAs that need to be fulfilled. \\ \hline
\textbf{Quality Attribute (QA)} & Characteristics that measure the quality of a system, such as performance, reliability, security, etc. In ARLO, QAs serve as the bridge between requirements and architectural choices. \\ \hline
\textbf{Condition Group (CG)} & A group of equivalent conditions from the lingual perspective. \\ \hline
\textbf{Concurrent Condition Group (CCG)} & A collection of CGs that can be true simultaneously. \\ \hline
\textbf{Decision Group} & A category of architectural decisions, such as deployment strategy, data caching method, etc. Each group contains multiple choices. \\ \hline
\textbf{Choice} & A specific architectural option within a Decision Group, such as "Monolith" or "Microservices" within the "Deployment" Decision Group. \\ \hline
\textbf{Decision} & A choice selected within a decision group. \\ \hline
\end{tabularx}
\end{footnotesize}
\end{table}

Before detailing ARLO's major steps and implementation details, we introduce a running example to illustrate the discussion.
\looseness-1
To this end, we use Urban Messaging System~(UMS), designed for high-density areas and capable of adapting between normal and disaster-response modes. UMS was chosen to demonstrate ARLO's ability to adapt to different operational conditions, each with specific desired QAs. 

Table~\ref{tab-messaging-reqs} lists ten requirements for UMS, with corresponding relevant {QA}s. Some requirements (R4, R5, and R10) may not imply any QAs; in other words, ARLO should not detect them as ASRs. On the other hand, multiple ASRs may target the same QA, impacting its weight. Among UMS's ten requirements, six are identified as ASRs and mapped to two conditions groups (right-most column)---normal operation (1)  and infrastructure failure (2)---as elaborated below.

\subsection{ARLO's Step 1: Extraction of ASRs Using LLM}\label{approach-step1}
We use the LLM with Prompt 1 from Table~\ref{tab:prompts} to extract a list of ASRs with associated QAs and, if applicable, any conditional statements. Each prompt in Table~\ref{tab:prompts} is designed to clearly define the LLM's expected task based on the provided data, followed by an explanation of specialized concepts, such as the definition of ASRs or a list of QAs and their meanings. This structured prompting aligns with approaches demonstrated to be effective in instruction-tuning LLMs~\cite{wei2021finetuned}.

\begin{table}[t!]
\begin{footnotesize}
\centering
\caption{UMS requirements and ARLO's outputs.}
\label{tab-messaging-reqs}
\begin{tabularx}{\columnwidth}{|p{0.25cm}|X|p{0.5cm}|p{1.35cm}|p{0.375cm}|}
\hline
\textbf{ID} & \textbf{Description} & \textbf{ASR} & \textbf{QA} & \textbf{CG} \\ 
\hline
R1 & System must support sub-second message delivery under normal operating conditions. & Y & Performance & 1 \\ 
\hline
R2 & During infrastructure failure, the system must automatically utilize neighboring devices to ensure delivery. & Y & Reliability & 2 \\ 
\hline
R3 & System should prioritize message delivery based on urgency during an emergency, as determined by embedded NLP algorithms. & Y & Usability, Reliability & 2 \\ 
\hline
R4 & System should offer both biometrics and social authentication. & N & N/A & - \\ 
\hline
R5 & System updates should be deployable remotely without interrupting ongoing messaging services. & N & N/A  & - \\ 
\hline
R6 & During a disaster, the system guarantees eventual message delivery without a specific time constraint. & Y & Reliability & 2 \\ 
\hline
R7 & In normal conditions, the system must encrypt messages end-to-end to ensure privacy and security. & Y & Security & 1 \\ 
\hline
R8 & It should integrate with existing social  platforms for wider accessibility. & N & N/A & - \\ 
\hline
R9 & In case of a disaster, the system should  send alerts to all users. & Y & Usability, Reliability & 2 \\ 
\hline
R10 & It must allow messages up to 10K words during normal conditions and up to 100 words during emergencies. & N &  N/A  & - \\ 
\hline
\end{tabularx}
\end{footnotesize}
\end{table}

\begin{table}[b!]
\begin{footnotesize}
\centering
\caption{GPT Prompts used by ARLO}
\begin{tabular}{|>{\raggedright\arraybackslash}p{8.4cm}|}
\hline
\textbf{Prompt 1. Parsing Requirements} \\
\hline

I have provided a set of software requirements. I want you to extract the following information: 
\begin{itemize}
    \item Whether it is architecturally significant. A requirement is architecturally significant if it satisfies both of these conditions: 1) It explicitly states a key decision regarding high-level software architecture. 2) It specifies one or more quality attributes regarding software architecture: \textit{[List of QAs and their descriptions from Table~\ref{tab-25010-QAs} provided here]}
    \item Find the QAs mentioned in the list above.
    \item The condition that should be true when the QAs are expected.
\end{itemize} \\
\hline
\textbf{Prompt 2. Determining Condition Groups} \\
\hline
If the following conditions \textit{[List of conditions]} mean the same thing, one can infer another or be considered a subset of another, return `True.' Otherwise, return `False.'\\
\\
\hline
\textbf{Prompt 3. Determining Concurrent Condition Groups} \\
\hline

Organize the provided set of conditions into groups where conditions in the same group can be true simultaneously. Once grouped, simply return the IDs of the conditions in each group enclosed in parentheses.\\

\hline
\end{tabular}
\label{tab:prompts}
\end{footnotesize}
\end{table}

ARLO employs the eight QAs shown in Table~\ref{tab-25010-QAs}, the first seven of which are design-oriented QAs defined in the ISO/IEC 25010 standard that targets software product quality~\cite{ISO-25010:2011}. The standard also includes functional suitability and safety, which we have not included. Functional suitability does not directly relate to architectural choices but addresses the characteristics of application-specific functional requirements, while safety is process- and runtime-oriented as defined in the standard. As an added eighth QA, we have incorporated cost efficiency since architectural decisions often depend on cost considerations of project implementation, operation, and maintenance. 

While the list of QAs in Table~\ref{tab-25010-QAs} is broadly applicable, it is important to note that ARLO does not mandate or depend on the specific choice of QAs; the list in Table~\ref{tab-25010-QAs} can be augmented as appropriate. For example, ARLO could readily incorporate the subdimensions of the QAs from the ISO/IEC 25010 standard (such as recoverability, a subdimension of reliability, or modularity, a subdimension of maintainability) by updating Prompt 1. Similarly, although the motivation behind this work is to explore any zero-shot LLM, we have opted to employ GPT\-4o model APIs since GPT-4o is known for its proficiency in processing and interpreting complex language data~\cite{liu2023gpteval}.

\begin{table}[t!]
\begin{footnotesize}
\centering
\caption{QAs defined in ISO/IEC 25010, used by ARLO}
\label{tab-25010-QAs}
\begin{tabularx}{\columnwidth}{|p{1.675cm}|X|}
\hline
\textbf{QA} & \textbf{Description} \\ 
\hline
Performance Efficiency (PE) & Relates to the performance relative to the resources used under stated conditions. Sub-characteristics include time behavior, resource utilization, and capacity.\\
\hline
Compatibility (CO) & Assesses software's ability to co-exist with independent software in a common environment sharing resources. It includes interoperability, co-existence, and compliance. \\
\hline
Interaction Capability (IC) & Measures how easy and satisfying the software is. It covers appropriateness recognizability, learnability, operability, user error protection, user interface aesthetics, and accessibility.\\
\hline
Reliability (RE) & Measures the software's capacity to maintain its performance level under stated conditions for a stated period. It includes maturity, fault tolerance, and recoverability.\\
\hline
Security (SE) & Covers the software's ability to protect information and data, ensuring confidentiality, integrity, non-repudiation, accountability, and authenticity. \\
\hline
Maintainability (MA) & Measures how easy it is to modify the software. It includes modularity, reusability, analyzability, modifiability, and testability.\\
\hline
Flexibility (FL) & Measures the ease with which the software can be transferred from one environment to another. It includes adaptability, installability, replaceability, and flexibility compliance.\\
\hline
Cost Efficiency (CE) & Emphasizes minimizing financial resources in software development, maintenance, and operation to stay within budget.\\
\hline
\end{tabularx}
\end{footnotesize}
\end{table}

\begin{figure}[b!] 
\vspace{-2mm}
  \footnotesize 
\begin{lstlisting}[language=json]
[ {
    "Id": 1,
    "IsArchitecturallySignificant": true,
    "QualityAttributes": ["Performance Efficiency"],
    "ConditionText": "under normal operating conditions" }, 
  {
    "Id": 2,
    "IsArchitecturallySignificant": true,
    "QualityAttributes": ["Reliability"],
    "ConditionText": "In the event of infrastructure failure" }, 
  {
    "Id": 3,
    "IsArchitecturallySignificant": true,
    "QualityAttributes": ["Usability", "Reliability"],
    "ConditionText": "during an emergency" },
  {
    "Id": 4,
    "IsArchitecturallySignificant": false,
    "QualityAttributes": [],
    "ConditionText": "N/A" },
 ... 
]
\end{lstlisting}
  \vspace{-2mm}
  \caption{\footnotesize An excerpt from ARLO's Step 1 Output for UMS}
  \label{fig:UMS-step1}
\end{figure}

As shown in Figure~\ref{fig:UMS-step1}, the output of this step of ARLO is a set of requirements classified as ASRs, each associated with specific QAs and possibly conditions under which the requirement occurs. If an ASR is unrelated to any condition (``N/A'' in Table~\ref{tab-messaging-reqs}), ARLO implicitly assigns it the default condition ``under any circumstances''. For instance, the first requirement in Table~\ref{tab-messaging-reqs} is correctly classified as ASR: it implies performance efficiency as a QA since it demands a sub-second message delivery. ARLO also correctly identifies its condition as ``under normal operating conditions'' as shown in Figure~\ref{fig:UMS-step1}. This step shapes subsequent architectural decision-making, as detailed below.

\subsection{ARLO's Step 2: Traceability by Grouping Desired QAs}\label{approach-step2}
\begin{figure}[ht!]
\footnotesize
\centering
\includegraphics[width=0.45\textwidth]{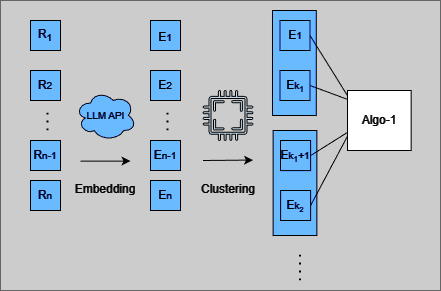}
\caption{Clustering ASRs before Applying Algorithm~\ref{algo-concern-formation}}
\label{fig:Algo1Optim}
\end{figure}

Once the ASRs, their corresponding QAs, and the conditions under which they hold are obtained, they are grouped based on the conditions so that ARLO can select a set of architectural choices for each group.
As illustrated in Figure~\ref{fig:Algo1Optim}, ARLO first uses LLM APIs \cite{openai_embeddings} to generate embeddings for each requirement. These embeddings serve as input to hierarchical clustering~\cite{hastie2009elements}, which groups requirements based on similar conditions. Notably, clustering is employed to minimize the number of LLM calls, which is the most time-intensive part of ARLO’s process, as shown later in this section. Hierarchical clustering is particularly well-suited here because it does not require a predetermined number of clusters, allowing for flexibility in grouping based on a varying number of conditions.

\begin{algorithm}
\begin{footnotesize}    
\caption{Categorize ASRs into Condition Groups}
\label{algo-concern-formation}
\begin{algorithmic}[1]
\STATE \textbf{Input:} List of ASRs within each cluster.
\STATE \textbf{Output:} List of condition groups within each cluster.
\STATE $groups \gets [~]$
\FORALL{$asr$ in $asrs$}
    \STATE $cond \gets extract\_cond(asr)$
    \STATE $group\_found \gets \text{false}$
    \FORALL{$group$ in $groups$} 
        \IF{logic\_equiv($cond$, $group$['nominal\_cond'])}
            \STATE $group$[`asrs'].append($asr$)
            \STATE $group\_found \gets \text{true}$
            \STATE $break$ 
        \ENDIF
    \ENDFOR
    \IF{not $group\_found$}
        \STATE $new\_group \gets \{nominal\_cond$: $cond$, asrs: [$asr$]\}
        \STATE $groups$.append($new\_group$)
    \ENDIF
\ENDFOR
\end{algorithmic}
\end{footnotesize}
\end{algorithm}

ARLO runs Algorithm~\ref{algo-concern-formation} within each cluster with Prompt~2 from Table~\ref{tab:prompts} to determine the condition groups (CGs). The algorithm creates a list variable called \textit{groups} (line 1). The algorithm forms a new group in the first iteration. It adds it to \textit{groups} since no existing condition group exists. In subsequent iterations, it tries to add more ASRs to an existing group or add a new group if needed. For each group, we keep one of the conditions as the ``nominal condition''. Each ASR's condition is extracted and compared with the nominal conditions of existing groups (lines 2-6). If the condition of the ASR is logically equivalent to the nominal condition of an existing group (based on LLM's determination using prompt 2 in Table~\ref{tab:prompts}), the ASR is added to that group (lines 7-10). If there is no logical equivalence with any existing group, a new group is formed with this new condition as its nominal condition, and the ASR is added to this group (lines 12-15). It is possible for a requirement to contain multiple conditions, meaning it could belong to more than one CG. Although the current implementation of Algorithm~\ref{algo-concern-formation} does not handle this scenario, a careful tracking mechanism will be necessary to avoid double-counting the implied QAs.

Note that Algorithm~\ref{algo-concern-formation} can be used directly to form CGs. However, doing so requires $O(n^2)$ pairwise comparisons, invoking the LLM API, which takes $l$ time units (with $l~\approx~100$ms/token for GPT-4o). The total time would thus be $O(n^2 \cdot l)$. To reduce API calls, we apply clustering, which is CPU-bound with complexity $O(n^3)$~\cite{wiki:hierarchical}. Once the clustering is done, we run Algorithm~\ref{algo-concern-formation} within each cluster. For $k$ clusters, each with $n/k$ requirements on average, the complexity becomes $O(n^2/k)$ per cluster. The overall complexity is, therefore,

\vspace{-3mm}
\begin{footnotesize}
\[
O(n^3) \, + O\left(\frac{n^2}{k} \cdot l \right) \, 
\]
\end{footnotesize}

In the above formula, the first term specifies the clustering complexity, and the second term specifies the reduced complexity in calling LLM APIs. The $l$ factor is orders of magnitude larger than the time required for CPU operations, thus reducing API calls shifts the main workload to local CPU resources. Clustering provides practical efficiency, especially for large-scale systems. However, it is important to note that this may slightly reduce the accuracy of generating CGs, as clustering based on embedding similarity may not be as precise as the lingual equivalency determined by LLMs \cite{freestone2024word}.

As an example of applying Algorithm~\ref{algo-concern-formation}, consider UMS requirements R1 and R6 from Table~\ref{tab-messaging-reqs}, both of which are ASRs. Initially, since there is no existing group, R1 is added to a newly created condition group CG1 with ``normal operating conditions'' as its nominal condition. When R2 is evaluated, its condition ``In the event of infrastructure failure'' is deemed not logically equivalent to the condition of the first group. Hence, it is added to a newly created condition group CG2. For each condition group, ARLO picks the QAs associated with the conditions included in that group. For example, since both R1 and R7 are members of CG1, their respective QAs, performance efficiency and security, are associated with CG1. Once the sets of CGs are determined, the LLM is used with Prompt 3 from Table~\ref{tab:prompts} to determine which CGs can be simultaneously true. Recall that we refer to such CGs as concurrent condition groups (CCGs). 

In the case of UMS, three nominal conditions are derived from ASRs: normal condition (NC), disaster response (DR), under any circumstances (UAC). We pass these nominal conditions as the input to LLM: 1) NC, 2) DR, and 3) UAC. We get (1, 3) and (2, 3) as the output, denoting pairs (NC, UAC) and (DR, UAC) can co-occur, thus forming two CCGs which we refer to as CCG1 and CCG2 subsequently. Note that it is impossible to have a case where both NC and DR are true simultaneously, while UAC always holds and can coexist with any other condition. 

In our approach, determining the weight of each QA is critical, as it reduces architects' bias and improves architecture alignment with ASRs. While stakeholders could directly provide these weights to reflect their priorities, ARLO derives these weights empirically based on the ASRs. Specifically, ARLO currently quantifies the weight of each QA by counting the number of ASRs that imply this particular QA. This can indicate the QA's priority as emphasized by several requirement. More sophisticated approaches could also be explored to determine QA priorities. For instance, analyzing the linguistic nuances within the requirements could reveal the criticality associated with a QA. For UMS, we ob the following weights: Performance: 1 (implied by R1), Reliability: 4 (implied by (R2, R3, R6, and R9), Usability: 2 (implied by R3 and R9), and Security: 1 (implied by R7). These weights are then used in the next step to guide the selection architectural choices.

\subsection{ARLO's Step 3: Architectural Choice Selection} \label{approach-step3}
In this step, the CCGs identified in the previous phase are leveraged to guide architectural decisions. ARLO employs a decision matrix to align the QAs with architectural decisions. In this matrix, rows represent architectural decision categories (e.g., deployment strategies, caching mechanisms), while columns represent the QAs. Note that architectural categories and choices used in rows are neither exhaustive nor prescriptive. They are illustrative examples intended to demonstrate how our approach can be applied and tailored to suit different projects or organizations' specific needs and contexts. Table~\ref{tab-matrix} shows a subset of the decision matrix relevant to the UMS case whose ASRs imply 4 QAs. (Section~\ref{sec:eval} shows cases where all QAs are implied.). The values shown in the table are based on the authors' experience; as previously mentioned, they can be customized to align with the unique requirements of specific projects. 

\begin{table}[htbp]
\begin{footnotesize}
\centering
\caption{QA-Arch Choice Mapping Matrix used with UMS}
\label{tab-matrix}
\begin{tabular}{|p{1.25cm}|p{2.3cm}|p{0.65cm}|p{0.65cm}|p{0.65cm}|p{0.65cm}|}
\hline
\textbf{Group} & \textbf{Choice} & \textbf{PE} & \textbf{IC} & \textbf{RE} & \textbf{SE} \\ 
\hline
\multirow{3}{=}{Deployment} & Monolith & 0 & 0 & -1 & 0 \tabularnewline
\cline{2-6}
 & Microservices & 1 & 0 & 0 & 0 \tabularnewline
\cline{2-6}
 & P2P & 0 & 0 & 1 & -1 \tabularnewline
\hline
\multirow{2}{=}{Data Caching} & Always-on & -1 & 1 & 1 & 1 \tabularnewline
\cline{2-6}
 & Offline First & 1 & 1 & 1 & -1 \tabularnewline
\hline
\multirow{3}{=}{Comm.} & API-Call & 0 & 0 & 0 & 1 \tabularnewline
\cline{2-6}
 & Message-Based & 1 & 0 & 1 & 0 \tabularnewline
\cline{2-6}
 & WebSocket & 1 & 1 & 0 & -1 \tabularnewline
\hline
\multirow{3}{=}{Data Repl.} & Central & 0 & 0 & -1 & 0 \tabularnewline
\cline{2-6}
 & Hot-Hot & 1 & 0 & 1 & 0 \tabularnewline
\cline{2-6}
 & Hot-Cold & 0 & 0 & 1 & 0 \tabularnewline
\hline
\multirow{3}{=}{Database Mgmt.} & SQL & 0 & 0 & 0 & 0 \tabularnewline
\cline{2-6}
 & NoSQL & 1 & 0 & 1 & 0 \tabularnewline
\cline{2-6}
 & Polyglot Persistence & 1 & 0 & 1 & 0 \tabularnewline
\hline
\multirow{2}{=}{Security} & Proactive & 1 & 0 & 0 & 1 \tabularnewline
\cline{2-6}
 & Reactive & 0 & 1 & 0 & 0 \tabularnewline
\hline
\multirow{2}{=}{Data Synch.} & Real-time  & 1 & 0 & 0 & 0 \tabularnewline
\cline{2-6}
 & Batch Processing & 0 & 1 & 1 & 0 \tabularnewline
\hline
\end{tabular}
\end{footnotesize}
\end{table}

The matrix values (+1), (0), or (-1) indicate whether a particular architectural choice supports, is neutral toward, or conflicts with a specific QA, respectively. An architect can specify these values based on their understanding of how decisions impact different QAs. For instance, based on the values we have chosen and shown in Table \ref{tab-matrix}, Microservices positively impact Performance Efficiency (+1). This positive impact is justified because Microservices enable independent scaling of different services based on their load, thereby optimizing resource allocation to achieve the desired performance by dedicating resources only where they are needed \cite{khazaei2020performance}. In contrast, Offline-first data caching can negatively impact Security (-1). This negative impact is justified because storing multiple copies of data locally increases the risk of exposure, heightening the chances of unauthorized access \cite{ali2021multi} or side-channel attacks \cite{liu2016newcache} and thereby, potential security breaches.

We then use integer linear programming (ILP) to maximize the QA satisfaction scores, thus selecting the optimal architectural choices. ILP aligns with the nature of our optimization goal in which each decision is binary and aims to optimize the total weighted sum of selections~\cite{ashouri2013optimal}. The optimizer considers the mutual exclusivity of choices within each group of choices. For example, ILP can select only one deployment option. Note that other approaches could be used instead of ILP. For instance, Dynamic Programming efficiently solves problems with overlapping sub-problems but requires substantial memory \cite{vadiyala2018exploring}. Genetic Algorithms are also suitable for exploring large search spaces without needing gradient information, but they do not guarantee a globally optimal solution~\cite{d2021gga}. 

Below is a formal definition of the optimization problem. The goal is to maximize the overall satisfaction score, defined as the sum of weighted satisfaction scores for each QA given a matrix $M$ with dimensions $n \times m$, each row represents an architectural choice, and each column represents a QA. The matrix element $M_{ij}$ denotes the effectiveness of choice $i$ in achieving QA $j$. Choices are grouped, and exactly one choice per group must be chosen. In the following formula \( x_i \) is a binary decision variable indicating the selection of choice \( i \) and \( W_j \) is the weights for QA\( j \):

\vspace{-2mm}
\begin{footnotesize}
\begin{align}
\text{Score}_j &= \sum_{i=1}^{n} M_{ij} \times x_i \label{eq:score} \\
\text{Maximize} &\sum_{j=1}^{m} \left( {Score}_j \times W_j \right) \text{ over } x_i \label{eq:maximization}
\end{align}
\end{footnotesize}

It is important to stress that the core of our work is not in the optimization algorithm itself but in how we model architectural decisions and formulate the objective function. The novelty lies in defining the decision matrix, which allows the problem to be treated as a linear optimization problem given by Equation~\ref{eq:maximization}. We use Google's linear optimization tool (OR-tools) \cite{or-tools} to determine the most suitable architectural choices based on the objective derived from the desired QAs and their weights for each condition group.  

Returning to the UMS case, we run ILP once for each condition group. The results of this process, reflecting the most effective alignment of architectural choices with the specified QAs under each condition group, are presented in Table \ref{table-messaging-ac}. These architectural choices are particularly tailored for scenarios such as disaster response, where the likelihood of a centralized infrastructure supporting other deployment strategies (e.g., microservices) is significantly reduced.

\begin{table}[t!]
\begin{footnotesize}
\centering
\caption{Choices for the Messaging System}
\label{table-messaging-ac}
\begin{tabular}{|p{2.5cm}|p{2.5cm}|p{2.5cm}|}
\hline
\textbf{Choice Group} & \textbf{CCG1} & \textbf{CCG2} \\ 
\hline
Deployment & Microservices & P2P \\ 
\hline
Data Caching & Always-on & Offline First \\ 
\hline
Communication & API-Call & Message-Based \\ 
\hline
Data Replication & Hot-Hot & Hot-Hot \\ 
\hline
DB Mgmt & NoSQL & NoSQL \\ 
\hline
Security Strategy & Proactive & Proactive \\ 
\hline
Data Synchronization & Real-Time & Batch Processing \\ 
\hline
\end{tabular}
\end{footnotesize}
\end{table}

\begin{table}[t!]
\begin{footnotesize}
\centering
\caption{QA Satisfaction Scores for the Messaging System}
\label{table-messaging-scores}
\begin{tabular}{|p{1.6cm}|p{3cm}|p{3cm}|}    
\hline
&\textbf{CCG1 (Normal)} &\textbf{CCG2 (Disaster)} \\ 
\hline
Performance & 4 & 0 \\ 
\hline
Reliability & 0 & 24 \\ 
\hline
Usability & 0 & 2 \\ 
\hline
Security & 3 & 0 \\ 
\hline

\end{tabular}
\end{footnotesize}
\vspace{-4mm}
\end{table}

Table \ref{table-messaging-scores} presents the satisfaction scores for QAs under each condition. Under normal conditions, the desired QAs are performance and security, with weighted scores of 4 and 3. The choice of a microservices deployment enhances performance, while an always-on data caching and API-call communication further contribute to this QA. The proactive strategy bolsters the system's security capabilities.

By contrast, operation during disaster response emphasizes reliability with a higher weighted score of 24 and interaction capability with a weighted score of 2. Note that these scores are calculated using Equation~\ref{eq:score}. This emphasis on reliability makes sense, given its higher weight. The optimal configuration for this group includes a P2P deployment, which significantly enhances reliability. Message-based communication and batch data synchronization also support this emphasis on reliability. An offline-first data caching strategy enhances Interaction Capability by allowing the system to function seamlessly even with intermittent or no internet connectivity. It achieves this by utilizing cached data and synchronization once connectivity is restored \cite{wang2021cost}.

This precise alignment of architectural choices with prioritized QAs mapped to various CCGs demonstrates the rigor of our method in architectural decision-making. It is important to note that CCGs are not mutually exclusive: a CG can belong to multiple CCGs. Each CCG provides a set of architectural decisions optimized for the nominal conditions of its associated CGs. The development team can review these CCGs and proactively implement the system to accommodate decisions from multiple CCGs. Furthermore, at runtime, a monitoring system can assess the current operational conditions and activate the appropriate architecture based on which CCG's conditions are met. 

\subsection{ARLO's Implementation}
ARLO has been implemented in C\#. The current prototype implementation consists of 1,626 source lines of code (SLOC). ARLO relies on two third-party packages: OpenAI's GPT-4 Completions API \cite{openai_api} and Google's linear optimization package \cite{google_ortools}. ARLO's complete source code is available at {\textit{https://anonymous.4open.science/r/ARLO-0BED}}.

\section{Evaluation}
\label{sec:eval}
To establish our approach's practical relevance, we evaluated three distinct cases, thoroughly examined in this section. The criteria for selecting these projects were full accessibility to requirements and a considerable ratio of ASRs (in our case, 15\% or higher). This threshold was chosen to ensure a substantial focus on architectural considerations.

With these cases, we aim to investigate the following research questions, with answers provided based on our findings from three case studies:

\begin{itemize}
    \item \textit{RQ1 -- How does ARLO scale when applied to software with varying numbers of requirements?} This RQ aims to determine ARLO's applicability to large, real-world systems.
    
    \item \textit{RQ2 -- How does change of requirements or the input matrix influence ARLO's outcomes?} This RQ's aim is to evaluate ARLO's sensitivity to requirements changes.
    
    \item \textit{RQ3 -- How do different conditions in requirements influence architectural choices?} This RQ aims to evaluate ARLO's tailorability to different requirements. 
\end{itemize}

\subsection{Dataset Description}
We selected three software systems---Atlassian Bamboo, Appcelerator Aptana, and Spring XD---with publicly available requirements~\cite{Choetkiertikul2018} to evaluate ARLO under different contexts and numbers of requirements. Table~\ref{tab-cases} summarizes the three systems' key statistics.


\begin{table}[b!]
\begin{footnotesize}
\centering
\caption{Case Statistics}
\label{tab-cases}
\begin{tabular}{|l|c|c|c|c|c|}
\hline
\textbf{} & \textbf{REs} & \textbf{Cleaned-Up} & \textbf{ASRs} & \textbf{Conditional} & \textbf{CCGs} \\ \hline
\textbf{Bamboo} & 522 & 374 & 94 & 84 & 11 \\ \hline
\textbf{Aptana} & 830 & 771 & 186 & 231 & 17 \\ \hline
\textbf{Spring XD} & 3759 & 3056 & 969 & 728 & 12 \\ \hline
\end{tabular}
\end{footnotesize}
\end{table}

As shown in Table~\ref{tab-cases}, the selected systems vary in size, with Bamboo having 522 requirements and Spring XD much larger with 3,759 requirements. Notably, in all three systems, at least 20\% of the requirements are identified as ASRs by ARLO and utilized to determine architectural choices.

\subsection{Experiment Results}
We executed ARLO's steps 1 and 2 as outlined in Section~\ref{sec:approach} to derive ASRs, identify the QAs associated with each ASR, group ASRs into CGs based on their conditions, and establish CCGs. As shown in Table~\ref{tab-cases}, each case contains several CCGs. Due to space constraints, we only show the results for the first CCG within each case, and the rest can be found in the appendix.\footnote{\url{https://anonymous.4open.science/r/ARLO-0BED/Appendix\%20B.pdf}} Table~\ref{tab-weights} displays the QA count for each case, revealing that some QAs are implied more frequently than others. ARLO selects the best options that satisfy the most frequently occurring QAs. This is a significant advantage of using ARLO, as human architects typically do not prioritize preferred QAs and instead aim to design an architecture that satisfies most QAs regardless of their priority \cite{kassab2015applying}.

\begin{table}[htbp]
\begin{footnotesize}
\centering
\caption{QA Counts for Three Cases}
\label{tab-weights}
\begin{tabular}{|l|c|c|c|c|c|c|c|c|}
\hline
\textbf{Case} & \textbf{PE} & \textbf{CO} & \textbf{IC} & \textbf{RE} & \textbf{SE} & \textbf{MA} & \textbf{FL} & \textbf{CE} \\ \hline
\textbf{Bamboo} & 8 & 7 & 15 & 13 & 9 & 16 & 4 & 2 \\ \hline
\textbf{Aptana} & 32 & 4 & 36 & 6 & 1 & 42 & 7 & 3 \\ \hline
\textbf{Spring XD} & 143 & 71 & 201 & 99 & 54 & 245 & 82 & 9 \\ \hline
\end{tabular}
\end{footnotesize}
\end{table}

Finally, we applied ARLO's step 3 to identify the architectural decisions for each case. Table~\ref{tab-results} shows the choices made for each case. As we can see, the results vary across cases. The main driver for the results is the QA counts given in Table~\ref{tab-weights}. For instance, we see that a monolith deployment has been chosen for both Spring XD and Aptana since they have a relatively high count for performance efficiency (PE). As we can see in the matrix, PE is positively linked with a monolith choice for deployment. This is not the case for Bamboo, which has a relatively low weight for PE.

\begin{table}[htbp]
\begin{footnotesize}
\centering
\caption{ARLO's Results for Three Cases}
\label{tab-results}
\begin{tabular}{|p{1.6cm}|c|c|c|}
\hline
\textbf{Group} & \textbf{Bamboo} & \textbf{Aptana} & \textbf{Spring XD} \\ \hline
Deployment & Microservices & Monolith & Monolith \\ \hline
Caching & Offline First & Offline First & Offline First \\ \hline
Communication & Message-Based & Message-Based & API-Call \\ \hline
Data Repl. & Hot-Cold & Central & Hot-Cold \\ \hline
DBMS & SQL & SQL & SQL \\ \hline
Security & Proactive & Proactive & Proactive \\ \hline
Data Synch. & Batch Processing & Real-time Sync & Real-time Sync \\ \hline
\end{tabular}
\end{footnotesize}
\end{table}

The significance of these findings lies in   ARLO's ability to adjust architectural decisions based on the specific QA profiles of each project. By leveraging the QA counts, ARLO ensures that the architecture is tailored to meet the most critical QAs. This contrasts with traditional manual approaches, where architects may overlook the nuanced importance of individual QAs, potentially leading to sub-optimal choices. 

\subsubsection{\textbf{RQ1 -- ARLO's Scalability}}
Our analytical treatment (provided in online appendix\footnote{\url{https://anonymous.4open.science/r/ARLO-0BED/Appendix\%20A.pdf}}) proves that selecting CGs is the most computationally intensive component within ARLO with an $O(n^2)$ complexity. Table~\ref{tab-scalability} shows ARLO's expected execution times for software with different numbers of requirements. In this table, the first column shows the number of requirements, the second column shows the iteration count required in Algorithm~\ref{algo-concern-formation}, and the last two columns show the best and worst-case running time. In the worst case, each requirement is an ASR and has a condition, and all conditions are mutually exclusive. This means the inner loop is Algorithm~\ref{algo-concern-formation} has to repeat equal to the number of $n$ groups added before it. So the total will be $0+1+...+n-1 = n(n-1)/2$. 

Based on the three cases presented earlier, the ASR ratio ranges from 18\% to 25\%, and the conditional requirements ratio ranges from 16\% to 27\%. Therefore, for the best-case scenario, we assumed a 15\% ASR ratio and a 15\% conditional ratio, which is strictly less than the range of values in the dataset. To calculate the times, we assumed it would take approximately 1s per LLM API call (20 tokens per requirement times 50ms per token ~\cite{OpenAICommunity2023}). The time is  calculated as the number of Iterations (column 2) times 1s for the worst case and scaled down by 0.0225 (15\% ASRs times 15\% conditionals) to obtain the best case values.

\begin{table}[t!]
\begin{footnotesize}
\centering
\caption{ARLO's Scalability}
\label{tab-scalability}
\begin{tabular}{|l|c|c|c|c|}
\hline
\textbf{System Size} & \textbf{\# Algo~\ref{algo-concern-formation} Iters} & \textbf{Best Case} & \textbf{Worst Case} \\ \hline
Small $\sim$ 100 REs & 4,950 & 10 Sec & 8 Min \\ \hline
Medium $\sim$ 1000 REs & 499,500 & 16 Min & 13 Hours \\ \hline
Large $\sim$ 2000 REs & 1,999,000 & 1 Hour & 55 Hours \\ \hline
X Large $\sim$ 5000 REs & 12,497,500 & 7 Hours & 14 Days \\ \hline
\end{tabular}
\end{footnotesize}
\end{table}

We ran ARLO with the matrix shown in Table~\ref{tab-casses-matrix}. We then measured the time needed to complete the process. The results align with the analytical expectations. ARLO processes Bamboo, which has 374 requirements (REs), in 10 minutes, and Aptana, with 771 REs, in 4 minutes. Both processing times fall within the expected range of 10 seconds to 16 minutes, as detailed in Table~\ref{tab-scalability}, which shows that it takes 10 seconds for 100 REs and 16 minutes for 1000 REs. Similarly, ARLO processes Spring XD, with 3056 REs, in 20 minutes, falling within the expected range of 16 minutes to 1 hour.

\begin{table}[htbp]
\begin{footnotesize}
\centering
\addtolength{\tabcolsep}{-0.5pt}
\caption{The Matrix Used to Experiment with Three Cases}
\label{tab-casses-matrix}
\begin{tabular}{|p{1.3cm}|l|p{0.24cm}|p{0.24cm}|p{0.24cm}|p{0.24cm}|p{0.24cm}|p{0.24cm}|p{0.24cm}|p{0.24cm}|}
\hline
\textbf{Group} & \textbf{Choice} & \textbf{PE} & \textbf{CO} & \textbf{IC} & \textbf{RE} & \textbf{SE} & \textbf{MA} & \textbf{FL} & \textbf{CE} \\ \hline
\multirow{3}{*}{Deployment} & Monolith & 1 & 0 & 1 & -1 & 1 & 0 & -1 & -1 \\ \cline{2-10} 
& Microservices & 0 & 1 & -1 & 0 & 0 & -1 & 1 & 0 \\ \cline{2-10} 
& Peer-to-Peer & -1 & -1 & 0 & 1 & -1 & 1 & 0 & 1 \\ \hline
\multirow{2}{*}{Caching} & Always-On & -1 & 0 & 1 & -1 & 1 & -1 & 0 & 1 \\ \cline{2-10} 
& Offline-First & 1 & 0 & -1 & 1 & -1 & 1 & 0 & -1 \\ \hline
\multirow{3}{*}{\parbox{1.8cm}{Communi-\\cation}} & API Calls & -1 & -1 & 1 & 0 & 1 & 1 & -1 & 0 \\ \cline{2-10} 
& Messaging & 0 & 0 & 0 & 1 & -1 & -1 & 1 & 0 \\ \cline{2-10} 
& Web-Socket & 1 & 1 & -1 & -1 & 0 & 0 & 0 & 0 \\ \hline
\multirow{3}{*}{\parbox{1.8cm}{Data\\Replication}} & Centralized & -1 & 0 & 0 & -1 & 0 & 1 & 0 & 1 \\ \cline{2-10} 
& Hot-Hot & 1 & 0 & 0 & 0 & 1 & -1 & 0 & -1 \\ \cline{2-10} 
& Hot-Cold & 0 & 0 & 0 & 1 & -1 & 0 & 0 & 0 \\ \hline
\multirow{2}{*}{DBMS} & SQL & -1 & 1 & 0 & -1 & -1 & 1 & 0 & 1 \\ \cline{2-10} 
& NoSQL & 1 & -1 & 0 & 1 & 1 & -1 & 0 & -1 \\ \hline
\multirow{2}{*}{Security} & Proactive & 0 & 0 & -1 & 0 & 1 & 1 & 0 & -1 \\ \cline{2-10} 
& Real-time & 0 & 0 & 1 & 0 & -1 & -1 & 0 & 1 \\ \hline
\multirow{2}{*}{\parbox{1.8cm}{Data\\Synch.}} & Reactive & 1 & 0 & 1 & -1 & 0 & -1 & 1 & -1 \\ \cline{2-10} 
& Batch & -1 & 0 & -1 & 1 & 0 & 1 & -1 & 1 \\ \hline
\end{tabular}
\end{footnotesize}
\end{table}

Based on our observations, ARLO can scale and adapt to increasing requirements, significantly reducing the time required compared to a manual approach. This scalability is particularly important for large and complex projects.

\subsubsection{\textbf{RQ2 -- ARLO's Sensitivity}}
RQ2 concerns what parameters used by ARLO have the most impact on its recommendations. Our experiments show that the two major factors impacting ARLO’s results are the matrix used and a particular set of requirements, which we call architecturally influential requirements (AIRs). While ASRs contain information that generally determine architectural decisions, AIRs are a subset that directly determines which ASRs, individually or in combination, can impact a specific decision. 

\textit{a) The matrix configuration}: If the matrix is configured to prefer choices equally, with the same sum across all rows for a given group, it won't impact ARLO's results, as the linear optimization treats all options equally. However, values set in the matrix can make a choice preferred by having a higher row sum. 
This indicates that architects using ARLO should carefully configure the matrix to obtain desirable results. If the matrix is unbalanced, they should ensure they can justify their preferences for any given choice.

\textit{b) Size of AIRs set}: AIRs can be key in determining ARLO's choices. AIRs imply QAs to which the optimizer's solution is the most sensitive. We can have cases where a single ASR forms an AIR or several ASRs form an AIR set. To select AIRs, 1) we first perform a sensitivity analysis to determine which QAs the results are most sensitive to. This involves applying deviations in QA counts ranging from 10\% to 90\%. The sensitivity is determined by observing how these adjustments affect the decisions. The QAs with most number of changes are selected as the most sensitive, 2) ASRs implying the most sensitive QA are removed individually. ARLO is then run to observe if the decisions change, 3) If the decisions change, an AIR set is identified, consisting of all the removed ASRs till the last decision change.

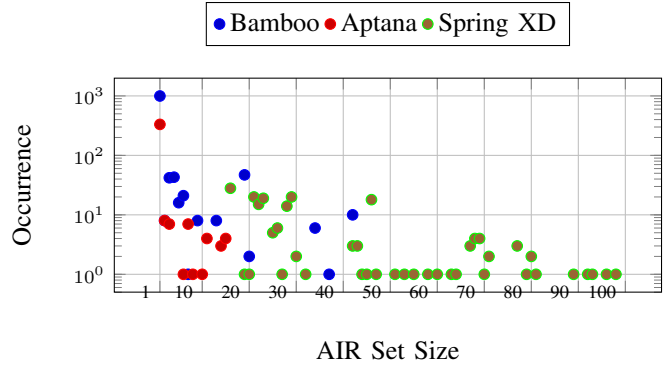
\begin{figure}[t!]
    \centering
    \begin{tikzpicture}
    \begin{axis}[
        width=1\columnwidth,
        height=0.5\columnwidth,
        xlabel={AIR Set Size},
        ylabel={Occurrence},
        xtick={1, 10, 20, 30, 40, 50, 60, 70, 80, 90, 100},
        xticklabels={1, 10, 20, 30, 40, 50, 60, 70, 80, 90, 100}, 
        ymode=log,
        log basis y={10},
        grid=major,
        legend style={at={(0.5,1.15)}, anchor=south,legend columns=-1}, 
        y tick label style={font=\scriptsize},
        x tick label style={font=\scriptsize, anchor=east} 
    ]

    \addplot+[only marks, mark=*, color=blue] coordinates {(1,997) (2,8) (3,42) (4,43) (5,16) (6,21) (7,1) (9,8) (13,8) (19,47) (20,2) (34,6) (37,1) (42,10)};
    
    \addplot+[only marks, mark=*, color=red] coordinates {(1,332) (2,8) (3,7) (6,1) (7,7) (8,1) (10,1) (11,4) (14,3) (15,4)};
    
    \addplot+[only marks, mark=*, color=green] coordinates {(16,28) (19,1) (20,1) (21,20) (22,15) (23,19) (25,5) (26,6) (27,1) (28,14) (29,20) (30,2) (32,1) (42,3) (43,3) (44,1) (45,1) (46,18) (47,1) (51,1) (53,1) (55,1) (58,1) (60,1) (63,1) (64,1) (67,3) (68,4) (69,4) (70,1) (71,2) (77,3) (79,1) (80,2) (81,1) (89,1) (92,1) (93,1) (96,1) (98,1)};
    
    \legend{Bamboo, Aptana, Spring XD}
    \end{axis}
    \end{tikzpicture}
    \caption{The AIR set size distribution}
    \label{fig:AIR-Set-Size-Dist}
\end{figure}

Figure~\ref{fig:AIR-Set-Size-Dist} shows that smaller systems like Bamboo have a higher frequency of smaller AIR sets, meaning that individual or small cluster requirements significantly influence the architecture. In contrast, larger systems like Spring XD tend to have AIR sets that encompass a broader range of requirements, reflecting the increased complexity of making the right architectural decisions in larger, more intricate systems. Both Bamboo and Aptana have a higher number of AIR sets consisting of a single requirement. In contrast, Spring XD's smallest AIR set consists of 15 ASRs, which is approximately 1\% of the total ASRs. This suggests that in large-scale projects, architectural decisions are less sensitive to individual requirements and more dependent on the collective influence of multiple interrelated requirements. 

\begin{table}[b]
\begin{footnotesize}
\centering
\caption{Smallest AIR set for Spring XD}
\label{tab-large-AIR-set}
\begin{tabular}{|>{\raggedright}p{0.5cm}|p{0.9cm}|p{6.2cm}|}
\hline
\textbf{ID} & \textbf{Values} & \textbf{Description} \\ \hline
R789 & RE, IC & Validate negative pageSize values in controllers. \\ \hline
R853 & PE, RE & Add connection pool to Redis connection factory. \\ \hline
R857 & IC, RE & RabitMQ sink module connection props override issue. \\ \hline
R859 & RE, IC & JDBC sink deletes table despite 'initDatabase=false'. \\ \hline
R898 & IC, RE & Shutdown servers using --cluster-name param. \\ \hline
R899 & IC, RE, PE & Use cleanup app from XD-1243 to stop previous CI runs on EC2. \\ \hline
R917 & IC, RE & Improve stream state management. \\ \hline
R924 & RE & Cannot access JMX endpoints with Boot Snapshot. \\ \hline
... & ... & ... \\ \hline
\end{tabular}
\end{footnotesize}
\end{table}

\textit{Impact analysis:}
We analyzed two single-AIR cases (Bamboo, Aptana) and one multi-AIR case (Spring XD) to assess ARLO’s sensitivity to requirements changes. Single-AIR cases are simpler to validate, while the multi-AIR case highlights ARLO’s ability to trace complex dependencies beyond manual methods. Note that the behavior of ARLO is independent of the requirement's complexity thus the chosen requirements generally represent the ASRs in the project.

\paragraph{\textit{Bamboo R127}: Reports indicate that Bamboo sometimes leaves 1orphaned' elastic instances and detached EBS volumes. Add functionality to allow admins to view and shut down instances not currently connected to Bamboo.}

Removing requirement R127, addressing reliability and cost efficiency, shifts from an `offline-first' data caching to`always-on' approach. R127 requires removing storage instances not connected to Bamboo, thus enhancing cost efficiency by reducing extra storage and ensuring reliability by avoiding outdated data. ARLO's decision to choose `always-on' can be explained as follows: In  Table~\ref{tab-casses-matrix}, `always-on' is less cost-efficient and less reliable than `offline-first' \cite{onyekwere2024, thenewstack2023}. By removing R127, the emphasis on cost efficiency and reliability is reduced. Thus, while both strategies are equal for reliability, `always-on', which is less cost-effective, is preferred.

Note that this transition would not have been easy for a human architect since the language of R127 focuses on a specific operational issue—removing orphaned storage instances. A human architect might address this requirement by implementing the specific functionality without considering the broader architectural implications. The requirement does not explicitly mention the `offline-first' or `always-on' strategies, so linking this operational fix to a change in the data caching approach requires a deeper analysis that is not immediately obvious.

\paragraph{\textit{Aptana R437}: Our current implementation of our JS parsing infrastructure does not allow the parser to be run outside of Eclipse. We need to extract the minimal set of classes that will de-couple our implementation from Eclipse.}          

Removing requirement R437, which addresses flexibility and maintainability, shifts from using SQL for DBMS to NoSQL. R437 recommends allowing Aptana to run outside of Eclipse, enhancing flexibility by decoupling it from Eclipse and improving maintainability by making it easier to run and debug from the command line. Removing R437 causes a transition to NoSQL because, in the Table~\ref{tab-casses-matrix}, Both SQL and NoSQL are equally preferred for flexibility, but SQL is favored for maintainability \cite{bharany2023comparative}. (An equal degree of flexibility is assumed because NoSQL offers greater adaptability to different types of data \cite{Roddam2023}, whereas SQL excels in terms of transferability across environments \cite{Wickramasinghe2023}. Since both adaptability and transferability are components of the flexibility definition provided in Table~\ref{tab-25010-QAs}, this results in an overall equal impact.) With R437 removed, the emphasis on both flexibility and maintainability is reduced. Therefore, there is less emphasis on maintainability, making NoSQL a viable option. 

Similar to the previous case, this transition would not have been easy for a human architect to arrive at. The language of R437 focuses on a specific technical implementation detail—decoupling the JS parser from Eclipse. A human architect might address this requirement by implementing the decoupling without considering the broader architectural implications. The requirement does not explicitly mention the impact on the database management system, so linking this technical change to a shift from SQL to NoSQL requires a deeper analysis that is not immediately apparent.

ARLO's approach systematically analyzes interdependencies across requirements and provides recommendations based on comprehensive data analysis, something that might be infeasible manually. The automated process can detect patterns and correlations that a human might overlook.

\paragraph{\textit{Spring XD AIR set}} As evident in Table~\ref{tab-weights}, the QA counts are significantly higher for Spring XD. This increase is due to the higher number of ASRs, implying more QAs. Consequently, the likelihood of having small AIR sets decreases. More ASRs must be removed to significantly change QA weights that can influence ARLO's decision-making process.

ARLO's ability to determine AIR sets in larger systems offers two key advantages for architects. First, identifying AIRs manually in smaller systems is feasible but impractical in larger ones. Table~\ref{tab-large-AIR-set} shows the smallest AIR set for Spring XD (only the first few requirements are described). Although not immediately apparent, ARLO identifies that these requirements all relate to reliability. Removing them could compromise reliability, so ARLO suggests transitioning from Hot-Cold to Centralized data replication. This demonstrates ARLO's ability to detect patterns and assist architects in making informed decisions. Second, ARLO limits the number of AIR sets, allowing for manageable reviews. Without focusing on AIRs, architects would need to consider all ASR combinations, which is impractical.

\subsubsection{\textbf{RQ3 -- Impact of Concurrent Condition Groups (CCGs)}}
Each CCG implies certain conditions that must be met for the CCG to occur when the system executes. Since each CCG includes QAs from different ASRs, we expect ARLO's results to vary across different CCGs. Our experiments confirm this variation, but the changes are insignificant due to the overlap among CCGs. Specifically, the ``under any circumstances'' group is included in all CCGs and typically has the most ASRs, greatly influencing QA counts. For example, in Aptana's CCG1 and CCG2, the choices for DBMS (SQL vs. NoSQL) and Security (Proactive vs. Reactive) differ.

This variation is due to different QA counts under each CCG, as shown in Table~\ref{tab-aptana-CCG-weights}. For instance, for CCG1, MA has the highest count, while IC has the highest count for CCG2. This results in different choices for DBMS, where MA demands SQL, while IC demands NoSQL. 

\begin{table}[htbp]
\begin{footnotesize}
\centering
\caption{QA Counts under Different CCGs for Aptana}
\label{tab-aptana-CCG-weights}
\begin{tabular}{|l|c|c|c|c|c|c|c|c|}
\hline
\textbf{CCG} & \textbf{MA} & \textbf{IC} & \textbf{PE} & \textbf{RE} & \textbf{CO} & \textbf{FL} & \textbf{CE} & \textbf{SE} \\ \hline
\textbf{CCG 1} & 46 & 41 & 41 & 11 & 8 & 8 & 3 & 2 \\ \hline
\textbf{CCG 2} & 43 & 45 & 40 & 16 & 7 & 7 & 3 & 1 \\ \hline
\end{tabular}
\end{footnotesize}
\end{table}

Specifying different architectural choices under various CCGs upfront allows the software to be implemented for specific conditions and transition to appropriate architectures based on operational conditions. 

\section{Limitations and Threats to Validity}\label{sec:ttv}

\textit{Dependence on LLM Accuracy}: ARLO relies heavily on LLMs for extracting ASRs and QAs. Any misinterpretation by the LLMs, especially in handling complex or ambiguous requirements, could lead to sub-optimal architectural decisions.

\subsection{Internal validity}
This threat involves potential biases in selecting architectural choices and QAs based on ISO/IEC 25010. To enhance validity, future work will involve broader expert input and using multiple LLMs to ensure consistency and reduce bias.

\subsection{Construct validity}
The configuration of the QA-architecture mapping matrix significantly shapes the  decisions ARLO recommends. An imbalanced matrix can introduce biases, making certain architectural choices appear more favorable. This requires careful calibration to ensure balanced and objective outcomes.

\subsection{External validity}
ARLO’s approach has been tested on specific software systems, which may not fully represent the diversity of all projects. Further research is needed to validate its applicability to other domains or more complex systems.

\section{Conclusion and Future Directions}
\label{sec:conclusion}
In this paper, we introduced ARLO, a novel approach that leverages LLMs and ILP to automate the translation of natural language requirements into software architectures. ARLO uses QAs as a bridge, enabling a systematic and traceable process for architectural decision-making. Our experiments demonstrated ARLO’s ability to scale with the number of requirements and highlighted its sensitivity to key architectural decisions. ARLO offers a flexible and data-driven tool to support decision-making processes for architects.

In future research, we would focus on refining the LLMs to identify architecturally significant requirements more accurately, deriving more sophisticated approaches to determine QA priorities, extending ARLO’s applicability to specific domains, exploring the impact of different matrix configurations on architectural outcomes, and handling overlaps among condition groups occurring due to multi-condition ASRs that could belong to multiple CGs. 

\bibliographystyle{IEEEtran}

\end{document}